\documentstyle[11pt]{article}
\begin{document}

\hskip 12cm
IPM-94-049
\vspace{1cm}

\begin{center}
{\Large \bf $Q$-Meromorphic Functions, Quantums sets and

Homomorphisms of the Quantum Plane}\\

\vspace{6mm}
{\bf Ahmad Shafei Deh Abad}\\
and\\
{\bf Vida Milani}\\
\vspace{6mm}
{\it Institute for Studies in Theoretical Physics and Mathematics(IPM),
Tehran,Iran.}\footnote{Mailing address:Institute for Studies in
Theoretical Physics and Mathematics(IPM),P.O.Box 19395-5746,Tehran,Iran.\\
E-mail:Shafei@netware2.ipm.ac.ir Milani@netware2.ipm.ac.ir}\\
{\it Department of Mathematics,Faculty of Science,University of
Tehran,Tehran,Iran.}\footnote{Permanent address}

\end{center}

\begin{abstract}

In this paper which is the completion of [1], we construct the $A_0(q)$-algebra
of $Q$-meromorphic functions on the quantum plane. This is the largest
non-commutative, associative, $A_0(q)$-algebra of functions constructed on the
quantum plane. We also define the notion of quantum subsets of R$^2$ which is
a generalization of the notion of quantum disc and charactrize some of their
properties. In the end we study the $Q$-homomorphisms of the quantum plane.

\end{abstract}

\newpage

{\bf 1 Introduction}\\

Non-commutative geometry and quantum groups are applied to problems of physics
in different ways. In particular classical and quantum mechanics on the
quantum plane have been studied in [2],[3] and [4]. As we have said in [1]
our main objection is to transfer classical mechanics on a Poisson algebra
${\cal B}$ to its functional quantization ${\cal A}$ in the sense of
[1]. More precisely we want to define an analogue of the Poisson bracket on
${\cal A}$ and to develope an appropriate classical mechanics on ${\cal A}$
parallel to that on ${\cal B}$. As a first step in this direction we gave a
new interpretation of the Manin quantum plane in [1]. As a result the class of
functions on the quantum plane was enlarged and the concept of $Q$-analytic
functions was introduced. In this paper which is a complement of [1] we try
to complete our necessary mathematical tools. It is worthwhile to note that
our approach results in some new mathematical constructions interesting in
themselves and which will be used in the formulation of classical mechanics
and studying the integrability and non-integrability of Hamiltonian
systems on quantum spaces.

In section 2 we enlarge the class of $Q$-analytic functions on the
quantum plane and construct the $A_0(q)$-algebra ${\cal M}_Q$ of
$Q$-meromorphic
functions on the quantum plane. This algebra in one hand contains the
$A_1(q)$-algebra of $Q$-analytic functions on the quantum plane as its
subalgebra, and on the other hand it can be considered as a quantization of
the C-algebra of the absolutely convergent power series
$\Sigma_{{i,j} >> - \infty} a_{ij}t_1^i t_2^j$ on R-\{0\}$\times$R-\{0\}.
The algebra ${\cal M}_Q$ appears to be non-commutative, associative
and having the unit element.
Let ${\cal s} :$R$^2$ $\rightarrow$R$^2$ be the permutation map.
It will be clear that if $f \in {\cal M}_Q$ then $fo{\cal s} \in {\cal M}_Q$,
and ${\cal M}_Q$ with
the above properties is maximal.

Section 3 is devoted to the study of the subsets of the quantum plane. In
[1] we introduced some very special subsets of the quantum plane, i.e, quantum
discs. Here we generalize this notion and define the quantum subsets of the
quantum plane. A quantum subset $\Omega$ is identified by the algebra of
$Q$-analytic functions on it. This algebra which is a non-commutative,
associative
$A_1(q)$-algebra is in fact a functional quantization of a certain subalgebra
of the commutative algebra of functions on $\Omega$. If for a subset of R$^2$
this subalgebra exists, then its functional quantization exists and is unique.
The properties of the $Q$-analytic functions helps us to find some
conditions for a subset of R$^2$ to be a quantum set. Some of the properties of
the quantum sets are also studied in this section. The algebra of functions
on these quantum subsets is larger than that of the quantum plane and so the
Hamiltonian systems on the quantum subsets has more chance to be integrable.
This problem will also be true for the case of R-\{0\}$\times$R-\{0\}.

In section 4 the morphisms of the quantum plane are investigated. In fact
these morphisms come from the homomorphisms of the $A_1(q)$-algebra
of $Q$-analytic
functions on the quantum plane, exactly in the same way that smooth mappings
of a manifold $M$ into itself come from endomorphisms
of $C^\infty (M)$ and vice versa.
So it is natural to consider the homomorphisms of the $A_1(q)$-algebra of
$Q$-analytic functions on the quantum plane instead of its morphisms
and y abuse of the language the homomorphisms of
${\cal A}_Q$ will be called the homomorphisms of the quantum plane. Manin
in his definition of $GL_q(2)$ uses the $2 \times 2$ matrices
$\left( \begin{array}{cc} a & b \\ c & d \end{array} \right).$
All one knows about these matrices is that their elements satisfy some
certain commutation relations. In his approach these elements commute with the
coordinate functions of the quantum plane $x$ and $p$. Our approach is
different, i.e., each automorphism of the quantum
plane is determined by a $2 \times 2$
matrix with elements in the $A_1(q)$-algebra of $Q$-analytic functions on the
quantum plane and so in general they do not commute with $x$ and $p$. The only
$Q$-analytic functions which commute with $x$ and $p$ are the elements
of $A_1(q)$
which commute with each other and so can not generate the
elements of the $GL_q(2)$ matrices.

Throughout this paper $t_1$ and $t_2$ will be the coordinate functions on the
ordinary plane R$^2$. The same functions considered as coordinate functions
on the quantum plane will be denoted respectively by $x$ and $p$ with the
well-known commutation relation $px = q xp.$ Moreover by the functional
quantization of a C-algebra we mean its maximal
$(1,D,A_1(q))$ or $(1,D-\{0\},A_0(q))$ functional
quantization in the sense of [1].
In the end it is worthwhile to say that the content of this paper and [1]
can be
generalized to any quantum space without any difficulty.\\

{\bf 2 $Q$-Meromorphic Functions}\\

Let D = \{q $\in$ C : $|q| \leq$1\} be the unit unit disc in C. As in [1],
$A_1(q)$ will be the C-algebra of all absolutely convergent power series
$\Sigma_{i=0}^\infty a_i q^i$ in D with values in C. Also we denote by
$A_0(q)$
the C-algebra of all absolutely  convergent power series
$\Sigma_{i>- \infty} c_i q^i$ on D-\{0\} with values in C. We can
generalize the
concept of $Q$-analytic functions on the 2-intervals of R$^2$ with values
in $A_1(q)$ to the algebra of $Q$-analytic functions on the
2-intervals of R$^2$ with values
in $A_0(q)$ without any difficulty. Assume now that
$\Omega = $R-\{0\} $\times$ R-\{0\} and let
$$ f=\Sigma_{i> - \infty} \Sigma_{j,k >> - \infty}a_{ijk} q^i t_1^j t_2^k$$
be an absolutely convergent power series on D-\{0\} $\times \Omega$ with
values in C.(The sign $>>$ under the second $\Sigma$ indicates that $j,k$ are
bounded below). Clearly we can consider $f$ as a function from $\Omega$
into $A_0(q)$ admitting the absolutely convergent Laurent expansion
$$f=\Sigma_{i,j>>- \infty}a_{ij}(q) t_1^i t_2^j$$ on $\Omega$. Since the above
series is absolutely convergent on $\Omega$, we can also write it as
$$f={\bar \Sigma}_{i,j=0}^{\infty} t_1^{-i} \alpha_{ij}(t_1,t_2) t_2^{-j}$$
where the $\alpha_{ij}$s are absolutely convergent power series on R$^2$
with values in $A_0(q)$ and the sign - over the $\Sigma$ means that the
indices are bounded above.

{\bf Definition 2.1.} With the above notations and conventions let
$${\hat f} = {\bar \Sigma}_{i,j=0}^{\infty} x^{-i}
{\hat \alpha_{ij}}(x,p) p^{-j}$$
be obtained from $f$ by the correspondence
$$t_1^i t_2^j = t_2^j t_1^i \rightarrow x^i p^j.\hspace {3cm} (+) $$
We call ${\hat f}$ a {\it Q-meromorphic} {\it function} on $\Omega$ with values
in $A_0(q)$ or simply a $Q$-meromorphic function on $\Omega.$

The two functions $\frac{1}{x}$ and $\frac{1}{p}$ are $Q$-meromorphic functions
on $\Omega$ satisfying the following commutation relations
$$x \frac{1}{x}=\frac{1}{x}x=1,p \frac{1}{p}=\frac{1}{p}p=1$$
$$p \frac{1}{x}=q^{-1} \frac{1}{x}p , x \frac{1}{p}=q \frac{1}{p}x$$
$$\frac{1}{p}.\frac{1}{x}=q \frac{1}{x}.\frac{1}{p},
{(\frac{1}{x})}^i=\frac{1}{x^i},{(\frac{1}{p})}^j=\frac{1}{p^j}.$$
By using these commutation relations we always follow the order
${(x^ip^j)}_{i,j>>- \infty}$
in writing the $Q$-meromorphic functions as above.

{\bf Remark.} If ${\hat f}(x,p)$ is a $Q$-analytic function
on the quantum plane
with values in $A_0(q),$ then for $k,l \in$ Z, ${\hat f}(q^kx,q^lp)$ is a
$Q$-analytic function on the quantum plane with values in $A_0(q).$

{\bf Definition 2.2.} The product of two $Q$-meromorphic functions
$${\hat f}={\bar \Sigma}_{i_1,i_2=0}^{\infty}
x^{-i_1}{\hat a}_{i_1i_2}(x,p)p^{-i_2}$$
$${\hat g}={\bar \Sigma}_{j_1,j_2=0}^{\infty} x^{-j_1}
{\hat b}_{j_1j_2}(x,p)p^{-j_2}$$
on $\Omega$ will be defined by
$${\hat f}.{\hat g}={\bar \Sigma}_{i_1,i_2=0}^{\infty}
{\bar \Sigma}_{j_1,j_2=0}^{\infty}
q^{i_2j_1}x^{-i_1-j_1}({\hat a}_{i_1i_2}(x,q^{-j_1}p).
 {\hat b}_{j_1j_2}(q^{-i_2}x,p))
p^{-i_2-j_2}$$
where the above product between ${\hat a}_{i_1i_2}$ and ${\hat b}_{j_1j_2}$ is
the product of two $Q$-analytic functions on the quantum plane with
values in $A_0(q)$ in the
sense of [1].

{\bf Lemma 2.1.} With the above notations the product of two
$Q$-meromorphic functions
${\hat f}$ and ${\hat g}$ on $\Omega$ is a $Q$-meromorphic
function on $\Omega.$

{\bf Proof.} The proof is easily seen from the fact that
${\hat a}_{i_1i_2}(x,q^{-j_1}p).
{\hat b}_{j_1j_2}(q^{-i_2}x,p)$ is a $Q$-analytic
function on the quantum plane with values
in $A_0(q).$

{}From the above lemma we can see that the set of all
$Q$-meromorphic functions on
$\Omega$ with values in $A_0(q)$ is a non-commutative, associative
$A_0(q)$-algebra
${\cal M}_Q$ with unity. This algebra contains ${\cal A}_Q$,
the $A_1(q)$-algebra
of $Q$-analytic functions on the quantum plane with values in
$A_1(q)$, as its
subalgebra. It is clear that ${\cal M}_Q$ is the
$(1,D-{0},A_0(q))$ functional quantization
of ${\cal M}$: the C-algebra of all absolutely convergent
power series $\Sigma_{i,j>>-\infty}a_{ij}t_1^it_2^j$
on $\Omega$ with values in C, and if we denote by ${\cal A}$ the
C-algebra of all
entire functions of the form $\Sigma_{i,j=0}^{\infty}a_{ij}t_1^i t_2^j$
on R$^2$ with values in C, then the following diagram commutes
$$\Phi_{\cal A}: {\cal A}_Q \rightarrow {\cal A}$$
$$\hspace{8mm} \downarrow \hspace{10mm} \downarrow$$
$$\Phi_{\cal M}: {\cal M}_Q \rightarrow {\cal M}$$
where $\Phi_{\cal A}$ and $\Phi_{\cal M}$ are the uantization
maps defined in [1]
and ${\cal A}_Q \rightarrow {\cal M}_Q$
and ${\cal A} \rightarrow {\cal M}$ are the canonical injections.\\

{\bf 3 Quantum Sets}\\

Let D be the unit disc in C introduced in section 2, and let $\Omega$ be a
non-empty subset of R$^2.$ Assume that
$$f=\Sigma_{i,j,k=0}^\infty a_{ijk}q^i t_1^j t_2^k$$
is an absolutely convergent power series on D$\times \Omega$ with values in C.
We can consider $f$ as a function on $\Omega$ with values in $A_1(q)$ admitting
the absolutely convergent Taylor expansion
$f=\Sigma_{i,j=0}^\infty a_{ij}(q)t_1^i t_2^j$
around (0,0) on $\Omega.$ Let ${\hat f}= \Sigma_{i,j=0}^ \infty a_{ij}x^i p^j$
be obtained from $f$ by the correspondence (+) in section 2.

{\bf Definition 3.1.} With the above notations ${\hat f}$
is called a {\it Q-analytic}
{\it function} {\it on} {\it the} {\it set} $\Omega$ with values in $A_1(q),$
or simply a $Q$-analytic function on $\Omega.$

Let $\Omega \subseteq$R$^2$ and ${\cal A}(\Omega)$ be the
maximal C-algebra consisting
of a) the C-algebra of all absolutely convergent power series
$\Sigma_{i,j=0}^\infty
a_{ij}t_1^it_2^j$ on R$^2$ and b) for each $(\alpha,\beta) \in$ R$^2$-$\Omega$
at least one analytic function on $\Omega$ admitting the
absolutely convergent Taylor
expansion $\Sigma_{i,j=0}^\infty a_{ij}t_1^it_2^j$ on $\Omega$ with values in C
having $(\alpha,\beta)$ as a singular point. Clearly ${\cal A}(\Omega)$ is a
commutative, associative, C-algebra with unity.

Let $\Omega$ be a subset of R$^2$ with non-empty interior, admitting the
maximal C-algebra
${\cal A}(\Omega)$ with the above properties. Then the
$(1,D,A_1(q))$ functional quantization
of this algebra is seen to be the maximal $A_1(q)$-algebra
consisting of a) $Q$-analytic functions on the quantum plane
with values in $A_1(q)$ and
b) for each $(\alpha,\beta) \in$R$^2$-$\Omega$ at least one $Q$-analytic
function on $\Omega$ with values in $A_1(q)$ which is not
defined at $(\alpha,\beta). $
This functional quantization is denoted by ${\cal A}_Q(\Omega)$ and is
clearly an associative, non-commutative, $A_1(q)$-algebra with unity.
It is also clear
that if for $\Omega \subseteq$R$^2$, ${\cal A}(\Omega)$ exists then
${\cal A}_Q(\Omega)$
exists and is unique.

{\bf Definition 3.2.} With the above notations and conventions,
${\cal A}_Q(\Omega)$
is called {\it the} {\it algebra} {\it of} {\it Q-analytic}
{\it functions} {\it on}
{\it the} {\it set} $\Omega$ and the pair $(\Omega,{\cal A}_Q(\Omega))$ or
simply $\Omega$ is called
a {\it quantum} {\it subset} of R$^2$ or simply a {\it quantum} {\it set}.

The following examples show that there exist some subsets of R$^2$ which are
quantum sets. The first example indicates that the definition of
the quantum set is
in fact a generalization of the definition of the quantum disc defined in [1].

{\bf Example 3.1.} let $\Omega \subseteq$R$^2$ be a disc with center (0,0).
We remind
from [1] that for each $(\alpha,\beta) \in$R$^2$ - $\Omega$, if we set
$$f_{\alpha,\beta}=\frac{1}{t_1^2+t_2^2-(\alpha^2+\beta^2)}=
\Sigma_{i,j=0}^\infty a_{ij}t_1^i t_2^j,$$
then ${\hat f}_{\alpha,\beta}=\Sigma_{i,j=0}^\infty a_{ij}x^i p^j$
is a $Q$-analytic function on $\Omega$ and can not be defined at
$(\alpha,\beta).$
So there is a maximal $A_1(q)$-algebra ${\cal A}_Q(\Omega)$ containing the
$A_1(q)$-algebra
generated by ${\cal A}_Q$ and ${\hat f}_{\alpha,\beta}$ for each
$(\alpha,\beta)
\in$R$^2$-$\Omega.$ Therefore $\Omega$ is a quantum set.

{\bf Example 3.2.} Let $\Omega \subseteq$R$^2$ be the interior of a
closed curve
in R$^2$ given by the polynomial equation $f(t_1,t_2)=0.$ Assume
that $(0,0)\in \Omega^o$ (the interior of $\Omega$) and $\Omega$ is symmetric
with respect to the $t_1,t_2$ axis in R$^2.$
For each $(c,d) \in$R$^2$-$\Omega$ let $(\alpha,\beta)$ be the intersection of
$\delta \Omega$ with the line passing through (0,0) and $(c,d)$ and let
$r=\frac{\sqrt{\alpha^2+\beta^2}}
{\sqrt{c^2+d^2}}.$ Set
$$g(t_1,t_2)=\frac{1}{f(rt_1,rt_2)}=\Sigma_{i,j=0}^\infty a_{ij}t_1^it_2^j$$
then ${\hat g}(x,p)=\Sigma_{i,j=0}^\infty a_{ij}x^ip^j$ is a
$Q$-analytic function
on $\Omega$ and is not defined at $(c,d).$ It follows that
$\Omega$ is a quantum set.

The following examples will be used later in this section.

{\bf Example 3.3.} Let $\Omega \subseteq$R$^2$ be the
polygone obtained from the intersection
of the two bands $$|t_1+t_2| < 1, |t_1-t_2| <1.$$
For each $(\alpha,\beta) \in$R$^2$-$\Omega,$ let $(a,b)$ be the
intersection of
$\delta\Omega$ (the boundary of $\Omega$) with the line passing
through (0,0) and $(\alpha,\beta)$ and let
$r=\frac{\sqrt{\alpha^2+\beta^2}}{\sqrt{a^2+b^2}}.$ Now if
$$f_{\alpha,\beta}=\frac{1}{{(t_1+t_2)}^2-r^2}.\frac{1}{{(t_1-t_2)}^2-r^2}=
\Sigma_{i,j=0}^\infty a_{ij}t_1^it_2^j,$$
then ${\hat f}_{\alpha,\beta}(x,p)=\Sigma_{i,j=0}^\infty a_{ij}x^ip^j$
is a $Q$-analytic
function on $\Omega$ and is not defined at $(\alpha,\beta).$ So there
is a maximal
$A_1(q)$-algebra ${\cal A}_Q(\Omega)$ generated by ${\cal A}_Q$
 and ${\hat f}_{\alpha,\beta}$ for
$(\alpha,\beta)\in$R$^2$-$\Omega.$ So $\Omega$ is a quantum set.

{\bf Example 3.4.} Let $a \in$R and $\Omega =$R $\times (-a,a).$
Then for $(\alpha,\beta)
\in$R$^2$-$\Omega$ if $$f_{\alpha,\beta}=\frac{1}{t_2^2-\beta^2}=
\Sigma_{i=0}^\infty a_it_2^i$$
then ${\hat f}_{\alpha,\beta}=\Sigma_{i=0}^\infty a_ip^i$ is a
$Q$-analytic function
on $\Omega$ and is not defined at $(\alpha,\beta).$ So $\Omega$ is a quantum
set and we call it the {\it quantum}
{\it horizontal} {\it band}. In the same way we see that
$\Theta=(-a,a)\times $R
is a quantum set which is called the {\it quantum} {\it vertical} {\it band}.

{\bf Proposition 3.1.} If the intersection of a family of quantum sets
has a non-empty interior, then it is a quantum set.

{\bf Proof.} Let ${(\Omega_i,{\cal A}_Q(\Omega_i))}_{i\in I}$
be a family of quantum
sets. Then every $Q$-analytic function on
$\Omega_i$ can be considered as a $Q$-analytic function on the intersection
$\bigcap_{i\in I}\Omega_i.$ With this hypothesis let ${\cal A}$ be the
$A_1(q)$-algebra
generated by $\bigcup_{i\in I}{\cal A}_Q(\Omega_i).$
Let $(\alpha,\beta) \in$R$^2$-$
\bigcap_{i\in I}\Omega_i,$ then for at least one
$i\in I,$ $(\alpha,\beta)\in$R$^2$-$\Omega_i,$
and so there is a  maximal $A_1(q)$-algebra containing ${\cal A}$ and
consisting
of ${\cal A}_Q$ and for each $(\alpha,\beta) \in$R$^2$ -
$\bigcap_{i\in I}\Omega_i$ at least
one $Q$-analytic function on $\bigcap_{i\in I}\Omega_i$ with singularity at
$(\alpha,\beta).$
This maximal $A_1(q)$-algebra is the algebra of $Q$-analytic functions on
$\bigcap_{i\in I}\Omega_i,$ i.e, ${\cal A}_Q(\bigcap_{i\in I}\Omega_i).$
So the proof is complete.

{\bf Corollary.} The open or closed 2-intervals in R$^2$ are quantum sets.

{\bf Proposition 3.2.} The closure of a quantum set is a quantum set.

{\bf Proof}. Let $\Omega$ be a quantum set with non-empty boundary.
Let $(\alpha,\beta) \in$
R$^2$-${\bar \Omega}$ (where ${\bar \Omega}$ is the closure of $\Omega$). Let
$(c,d)$ be the intersection of $\delta \Omega$ with the line passing
through (0,0)
and $(\alpha,\beta).$ There exists a point $(a,b)\in$R$^2$-${\bar \Omega}$
on this
line between $(c,d)$ and $(\alpha,\beta)$ and a
$Q$-analytic function ${\hat f}$ on $\Omega$ which is not defined at
$(\alpha,\beta).$
Now let $r=\frac{\sqrt {a^2+b^2}}{\sqrt {\alpha^2+\beta^2}}$, then the
function ${\hat g}_{\alpha,\beta}(x,p)=
{\hat f}(rx,rp)$ is a $Q$-analytic
function on ${\bar \Omega}$ and is not defined at $(\alpha,\beta).$
So there exists a maximal
$A_1(q)$-algebra containing the $A_1(q)$-algebra
generated by ${\cal A}_Q$ and all the ${\hat g}_{\alpha,\beta}$s for each
$(\alpha,\beta)\in$R$^2$-${\bar \Omega}.$ this maximal algebra is the algebra
of $Q$-analytic functions on ${\bar \Omega}.$
So ${\bar \Omega}$ is a quantum set.

{\bf Lemma 3.1.} Let $\Omega \subseteq$R$^2$ and ${\hat f}$ be
a $Q$-analytic function
on $\Omega.$ Then for each $(\alpha,\beta) \in \Omega, {\hat f}$ is a
$Q$-analytic
function on the open open 2-interval $(-\alpha,\alpha) \times (-\beta,\beta)
\subseteq$R$^2.$

{\bf Proof.} Let ${\hat f}=\Sigma_{i,j=0}^\infty a_{ij}x^ip^j.$
Then the series
$\Sigma_{i,j=0}^\infty a_{ij}t_1^it_2^j$ is absolutely convergent
on the open 2-interval
$(-\alpha,\alpha) \times (-\beta,\beta) \subseteq$R$^2.$ The proof is complete.

{\bf Corollary.} If $\Omega \subseteq$R$^2$ is a quantum set, then for
$(\alpha,\beta) \in \Omega,$
the open 2-interval $(-\alpha,\alpha)\times(-\beta,\beta)$ is contained in
$\Omega.$ Consequently every quantum set contains the origin and
is symmetric with
respect to the coordinate axis.

{\bf Proposition 3.4.} Every convex set $\Omega \subseteq$R$^2$
containing (0,0) and
symmetric with respect to the coordinate axis is a quantum set if its
interior $\Omega^o$ is non-empty.

{\bf Proof.} If $\Omega=$R$^2,$ then clearly it is a quantum set.
So let $\Omega
\subset$R$^2.$ Since $\Omega$ is convex then for each
$(\alpha,\beta)\in$R$^2$-$\Omega$
there is a line passing through $(\alpha,\beta)$ and having empty intersection
with $\Omega.$ If $\Omega$ is not bounded since it contains the origin and
is symmetric with respect to the coordinate axis then for
each $(\alpha,\beta)\in$R$^2$-$\Omega$ there is a band
$(-\alpha,\alpha)\times$R
(or R$\times(-\beta,\beta)$) containing $\Omega.$ Let
$f=\frac{1}{t_1^2-\alpha^2}=\Sigma_{i=0}^\infty a_it_1^i$
(or $f=\frac{1}{t_2^2-\beta^2}=\Sigma_{i=0}^\infty a_it_2^i$),
then ${\hat f}_{\alpha,\beta}=
\Sigma_{i=0}^\infty a_ix^i$ (or ${\hat f}_{\alpha,\beta}=
\Sigma_{i=0}^\infty a_ip^i$) is a $Q$-analytic function on $\Omega$ and is
not defined at $(\alpha,\beta).$
[See example 3.4.]. If $\Omega$ is bounded then from the assumptions on
$\Omega$ it follows that for $(\alpha,\beta)\in$R$^2$-$\Omega$
there is a subset of R$^2$ with equation $|t_1+t_2|<r,|t_1-t_2|<r$ (or
$(-\alpha,\alpha)\times(-\beta,\beta)$) containing $\Omega$ and
having the point
$(\alpha,\beta)$ on its boundary. Let $$f=\frac{1}{{(t_1+t_2)}^2-r^2}.
\frac{1}{{(t_1-t_2)}^2-r^2}
=\Sigma_{i,j=0}^\infty a_{ij}t_1^it_2^j$$
(or $f=\frac{1}{t_1^2-\alpha^2}.\frac{1}{t_2^2-\beta^2}=
\Sigma_{i=0}^\infty a_{ij}t_1^it_2^j$),
then ${\hat f}_{\alpha,\beta}=\Sigma_{i,j=0}^\infty a_{ij}x^ip^j$ is
a $Q$-analytic
function on $\Omega$ and is not defined at $(\alpha,\beta)$
[see examples 3.3,3.4]. In
each case there is a maximal $A_1(q)$-algebra containing the
$A_1(q)$-algebra generated
by ${\cal A}_Q$ and ${\hat f}_{\alpha,\beta}$ for each
$(\alpha,\beta)\in$R$^2$-$\Omega.$
This maximal $A_1(q)$-algebra is the algebra of $Q$-analytic functions on
$\Omega.$
The proof is complete.

The following example shows that not all the quantum sets are convex.

{\bf Example 3.5} Let $\Omega \subseteq$R$^2$ be the intersection of
two sets given
by the equations $$t_1t_2<1,t_1t_2<-1,$$
then for each $(\alpha,\beta)\in$R$^2$-$\Omega$ let $(a,b)\in \delta \Omega$
be the intersection of $\delta \Omega$ with the line passing through (0,0)
and $(\alpha,\beta).$
Let $\Omega'$ be obtained from $\Omega$ by the coorespondence
$$\Omega \rightarrow \Omega'$$
$$(t_1,t_2) \mapsto \frac{\sqrt{a^2+b^2}}{\sqrt{\alpha^2+\beta^2}}(t_1,t_2).$$
Let $\lambda = \frac{\sqrt{\alpha^2+\beta^2}}{\sqrt{a^2+b^2}}$ and
$$f=\frac{1}{t_1t_2-\lambda^2}.
\frac{1}{t_1t_2+\lambda^2}=\Sigma_{i,j=0}^\infty a_{ij}t_1^it_2^j$$
then ${\hat f}_{\alpha,\beta}=\Sigma_{i,j=0}^\infty a_{ij}x^ip^j$ is a
$Q$-analytic
function on $\Omega$ and is not defined at $(\alpha,\beta).$
So there is a maximal $A_1(q)$-algebra containing the $A_1(q)$-algebra
generated by ${\cal A}_Q$
and ${\hat f}_{\alpha,\beta}$ for each $(\alpha,\beta)\in$R$^2$-$\Omega.$
This maximal algebra is the algebra of $Q$-analytic functions on $\Omega.$

{\bf Proposition 3.5.} Every quantum set $\Omega$ is contractible.

{\bf Proof.} From lemma 3.1 it is clear that
$$H : [0,1]\times \Omega \rightarrow \Omega$$
$$(t,\omega) \mapsto t\omega$$ for $t\in[0,1]$ and $\omega\in \Omega$
is the contraction map.\\

{\bf 4 Homomorphisms of the Quantum Plane}\\

{\bf Definition 4.1.} An $A_1(q)$-homomorphism
$\Theta : {\cal A}_Q \rightarrow {\cal A}_Q$
is called a {\it homomorphism} of the quantum plane.

{\bf Remark.} In order to determine a homomorphism $\Theta$ of
the quantum plane it is
sufficent to define $\Theta(x)$ and $\Theta(p)$ and then extend it by
$$ \Theta(\Sigma_{i,j=0}^\infty a_{ij}x^ip^j)=
\Sigma_{i,j=0}^\infty a_{ij} {\Theta(x)}^i{\Theta(p)}^j$$
to the whole ${\cal A}_Q.$ Now let $\Theta : {\cal A}_Q \rightarrow {\cal A}_Q$
be a homomorphism
of the quantum plane such that $\Theta(x)\neq 0$ and $\Theta(p)\neq 0.$ Since
$$\Theta(p) \Theta(x) = q \Theta(x) \Theta(p)$$
there are $Q$-analytic functions $f,g,h,k$ on the quantum plane such that
$$\Theta(x)=xf+gp$$ $$\Theta(p)=xh+kp$$
and the following relation satisfies:
$$x^2[h(x,qp)f(x,p)-qf(x,qp)h(x,p)]+$$
$$xp[h(q^{-1}x,p)g(q^{-1}x,p)-q^2g(q^{-1}x,qp)h(x,p)+qk(q^{-1}x,qp)f(x,p)-
qf(x,p)k(q^{-1}x,p)]+$$
$$p^2[k(q^{-2}x,p)g(q^{-1}x,p)-qg(q^{-2}x,p)k(q^{-1}x,p)]=0. \hspace{3cm} (*)$$
The set of all homomorphisms of the quantum plane is denoted by $Hom_Q$(R$^2$).

{\bf Definition 4.2.} The invertible elements of $Hom_Q$(R$^2$) are called
the {\it automorphisms}
of the quantum plane.

{\bf Remark.} If $\Theta : {\cal A}_Q \rightarrow {\cal A}_Q$ is an
automorphism of the
quantum plane and $$\Theta(x)=xf+gp,\Theta(p)=xh+kp$$
then there are $Q$-analytic functions $f_1,h_1$ on the quantum plane such
that we can write $$\Theta(x)=f_1x+gp,\Theta(p)=h_1x+kp.$$
So we can correspond a $2\times 2$ matrix to $\Theta$ by
$$\left( \begin{array}{cc} f_1 & g \\ h_1 & k \end{array} \right)
\left( \begin{array}{c} x \\ p \end{array}
\right)=\left( \begin{array}{c} \Theta(x) \\ \Theta(p) \end{array} \right).$$
The set of automorphisms of the quantum plane with the composition rule
is a group. But in the above matrix form if $\Theta_1,\Theta_2$ are two
automorphisms
of the quantum plane then the product of their corresponding matrices does not
necessarily correspond to an automorphism of the quantum plane. To see
this, let $$\left( \begin{array}{c} \Theta_1(x) \\ \Theta_1(p) \end{array}
\right)=\left( \begin{array}{cc} e^p & 0 \\ 0 & 1 \end{array} \right)
\left( \begin{array}{c} x \\ p \end{array} \right),$$
$$\left( \begin{array}{c} \Theta_2(x) \\ \Theta_2(p) \end{array} \right)=
\left( \begin{array}{cc} 1 & 0 \\ 0 & e^x \end{array} \right)
\left( \begin{array}{c} x \\ p \end{array} \right).$$
Then $$\left( \begin{array}{cc} e^p & 0 \\ 0 & 1 \end{array} \right) \left(
\begin{array}{cc} 1 & 0 \\ 0 & e^x \end{array} \right)
\left( \begin{array}{c} x \\ p \end{array} \right)=
\left( \begin{array}{cc} e^p & 0 \\ 0 & e^x
\end{array} \right) \left( \begin{array}{c} x \\ p \end{array}
\right).$$
But the elements of the product matrix do not satisfy $(*)$, since
$$e^{qp}.e^x \neq e^{q^{-1}x}.e^p.$$
But in the limit $q \rightarrow 1,$ the product of matrices corresponding to
the automorphisms will correspond to an automorphism. So clearly we can
consider the set of matrices corresponding to
the automorphisms of ${\cal A}_Q($R$^2$) as a functional
quantization of the qroup of automorphisms of ${\cal A}$(R$^2$) in the
sense that the elements of the matrices are functional quantizations of the
elements of the matrices
corresponding to the automorphisms of ${\cal A}$(R$^2$).

\vspace{1cm}
\begin{center}
{\bf References}
\end{center}

[1] A. Shafei Deh Abad and V. Milani, {\it J.M.P,vol.35,no.9,5074-5086.}

[2] S. V. Shabanov, {\it J.Phys.A:Math.Gen.26(1993)2583-2606.}

[3] I. Ya. Arefeva and I. V. Volovich, {\it Phys.letter B,268(1991)179-187.}

[4] J. Schwenk and J. Wess, {\it Phys.letter B,291(1992)273-277.}

\vspace{1cm}
\begin{center}
{\bf Acknowledgements}
\end{center}
The authors would like to thank Professor Abdus Salam, The International
Atomic Energy Agency and Prof.Narasimhan for hospitality at the
International Centre for Theoretical Physics,
Trieste,Italy where part of this work has been done.
\end{document}